\documentclass[aps,preprint,floatfix]{revtex4-2}
\pdfoutput=1
\usepackage{graphicx,bm,amsmath,dcolumn}
\RequirePackage{color}

\definecolor{MyDarkGreen}{rgb}{0.02,0.60,0.06}

\begin{document}
	
	\title{Quantum Magnetic Properties, Entanglement for Antiferromagnetic Spin 1 and 3/2 Cluster Models}
	
    \author{N.\ Ananikian}
	\email{ananik@yerphi.am}
    \affiliation{ AANL, Alikhanian Br. 2, 0036 Yerevan, Armenia}
	\affiliation{CANDLE, Acharyan 31, 0040 Yerevan, Armenia}
    \author{Vl.\ V.\ Papoyan}
	\email{vpap@theor.jinr.ru}
	\affiliation{BLTP, JINR, 141980 Dubna, Russian Federation}
	\affiliation{Dubna State University, Dubna, Russian Federation}

	\begin{abstract}

 Entanglement,  magnetization and magnetic susceptibility for 1D antiferromagnetic spin $1$ and spin $\frac{3}{2}$ Heisenberg $XXX$ model with Dzyaloshinskii-Moriya interaction, single-ion anisotropy and external magnetic  field on the finite chain are obtained.

	\end{abstract}

	\maketitle
	
\section{Introduction}
\label{Introduction}

Ferromagnetic and antiferromagnetic quantum magnetism has become an extremely active area of research \cite{1,4}.
Experimental data on the supramolecular architecture, synthesis, magnetic susceptibility with dinuclear, tetranuclear, heptanuclear, decanuclear, and polynuclear complexes of Cu$_{2}$ ions in the antiferromagnetic case are found \cite{5}. The preparation of an antiferromagnetic trinuclear Ni$_2$ ionic complex with spin $1$ and magnetic susceptibility via temperature have been obtained \cite{6}.


X-ray crystallographic structures, magnetic susceptibility from $2$ to $300 K$ and the theoretical analysis of magnetism for a triangular and a tetranuclear molecule consisting of linked high-spin cobalt (II) centers were described by the inclusion of an antisymmetric exchange term (Dzyaloshinsky-Moriya interaction) \cite{11,12}.

Thermal non-classical correlations quantified by concurrence entanglement, local quantum the uncertainty and quantum coherence in a four-qubit square chain (tetranuclear copper (II) complex) of mixed ferromagnetic-antiferromagnetic interactions between nearest neighboring spins with a magnetic field were examined exactly in \cite{13}. Magnetic susceptibility can reveal spin entanglement between individual constituents of a solid, and magnetization describes their local properties \cite{14}.

In this paper, we consider analytically the thermal entanglement,  magnetization and magnetic susceptibility of antiferromagnetic high-spin dinuclear and trinuclear Ni$_{2}$ (spin $1$) and Co$_{2}$ (spin $\frac{3}{2}$) metal ion complexes as a function of an external magnetic field and single-ion anisotropy at low temperatures.
\section{Model and Definitions}
The Hamiltonian for 1D antiferromagnetic  Heisenberg XXX model with Dzyaloshinskii-Moriya (DM) interaction and  single-ion anisotropy in external magnetic field given as
\begin{eqnarray}
   H & = & J \sum _{i=1}^N \vec{S}_{i}\vec{S}_{i+1} + G \sum _{i=1}^N (S_{i}^{x}S_{i+1}^{y}-S_{i}^{y}S_{i+1}^{x})+
   D \sum_{i=1}^N(S_i^{z})^2-B\sum _{i=1}^N S_i^z.  \label{ham}
\end{eqnarray}
where $J = 1$ to fix the energy scale, $G$ is DM interaction parameter, $D$ represents uniaxial single-ion anisotropy,
$B$ is the controlable parameter. The local spin vector $S_i$ for each site has the components of the spin-1 operators:
\begin{eqnarray}
&S_{x}&=\frac{1}{\sqrt{2}} \left(\nonumber
\begin{array}{lll}
 0 & 1 & 0 \\
 1 & 0 & 1 \\
 0 & 1 & 0
\end{array}
\right), \hskip 0.4cm
S_{y}=\frac{1}{\sqrt{2}} \left(
\begin{array}{lll}
 0 & -i & 0 \\
 i & 0 & -i \\
 0 & i & 0
\end{array}
\right), \hskip 0.4cm
S_{z}= \left(
\begin{array}{lll}
 1 & 0 &  0 \\
 0 & 0 &  0 \\
 0 & 0 & -1
\end{array}
\right).
\end{eqnarray}

The components of the spin $\frac{3}{2}$ operators are expressed as follows:
\begin{eqnarray}
S_{x}=\frac{1}{2}\left(\nonumber
\begin{array}{llll}
 0 & \sqrt{3} & 0 & 0 \\
 \sqrt{3} & 0 & 2 & 0 \\
 0 & 2 & 0 & \sqrt{3} \\
 0 & 0 & \sqrt{3} & 0 \\
\end{array}
\right)\hskip 0.4cm
S_{y}=\frac{1}{2}\left(
\begin{array}{llll}
 0 & \sqrt{3} & 0 & 0 \\
 \sqrt{3} & 0 & 2 & 0 \\
 0 & 2 & 0 & \sqrt{3} \\
 0 & 0 & \sqrt{3} & 0 \\
\end{array}
\right)\hskip 0.4cm
S_{z}=\frac{1}{2} \left(
\begin{array}{llll}
 3 & 0 & 0 & 0 \\
 0 & 1 & 0 & 0 \\
 0 & 0 & -1 & 0 \\
 0 & 0 & 0 & -3 \\
\end{array}
\right).
\end{eqnarray}

Logarithmic negativity \cite{15} is an entanglement measure defined as
\begin{eqnarray}
LN= \log_2\left(\parallel\rho_{ab}^{T_a}\parallel_{1}\right)=\log_2\left(\sum _{k=1}|\lambda_k|\right)
\label{ln}
\end{eqnarray}
where $\lambda_k$ are the eigenvalues of the matrix $\rho_{ab}^{T_a}$. $T_a$ denotes the partial pairwise transpose with respect to the first system and $\rho_{ab}$ is partial trace of the density matrix of the whole system between quantum states $a$ and $b$.

\section{Calculated Results}

\begin{figure}[!ht]
  	\includegraphics[width=120mm]{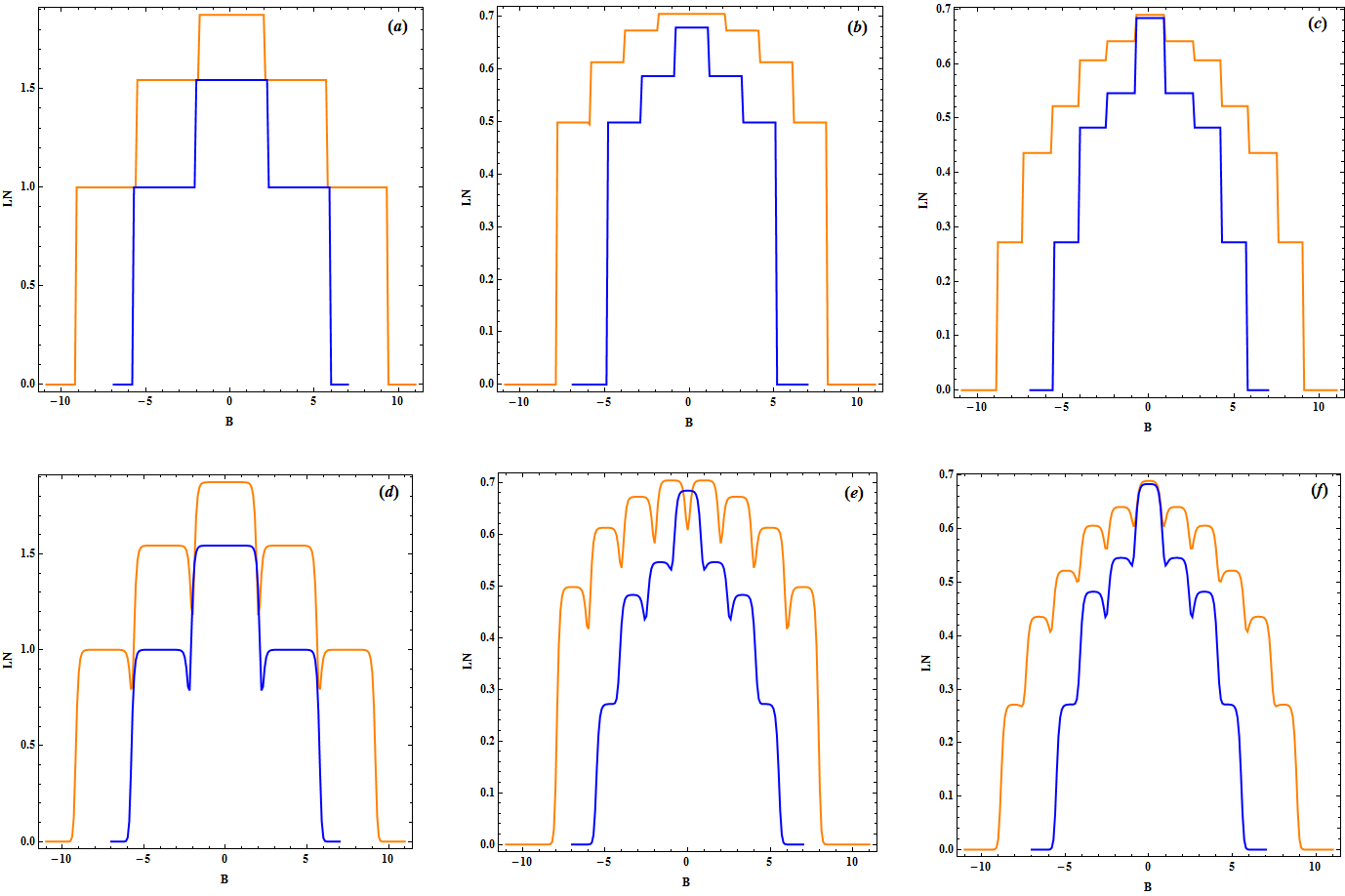}
	\caption{The logarithmic negativity $LN$ versus of external magnetic  field parameter $B$}
	\label{ba}
\end{figure}

The entanglement of the 1D antiferromagnetic spin $1$ and spin $\frac{3}{2}$ Heisenberg $XXX$ model with Dzyaloshinskii-Moriya interaction, single-ion anisotropy and an external magnetic  field on finite chain with sizes $N= 2$, $N = 3$ and $N = 4$ is studied.
The behavior of logarithmic negativity $LN$ as a function of the external magnetic  field parameter $B$ for spin $1$ and spin $\frac{3}{2}$ is displayed in Fig.\ \ref{ba}, for fixed $J = 1$, $D = 1$ and $G = 1$.
The critical point $B_c$ at $B \geq B_c$ of the ground state, at which entanglement disappears, is shown if $B > 0$.

\begin{figure}[!ht]
  	\includegraphics[width=120mm]{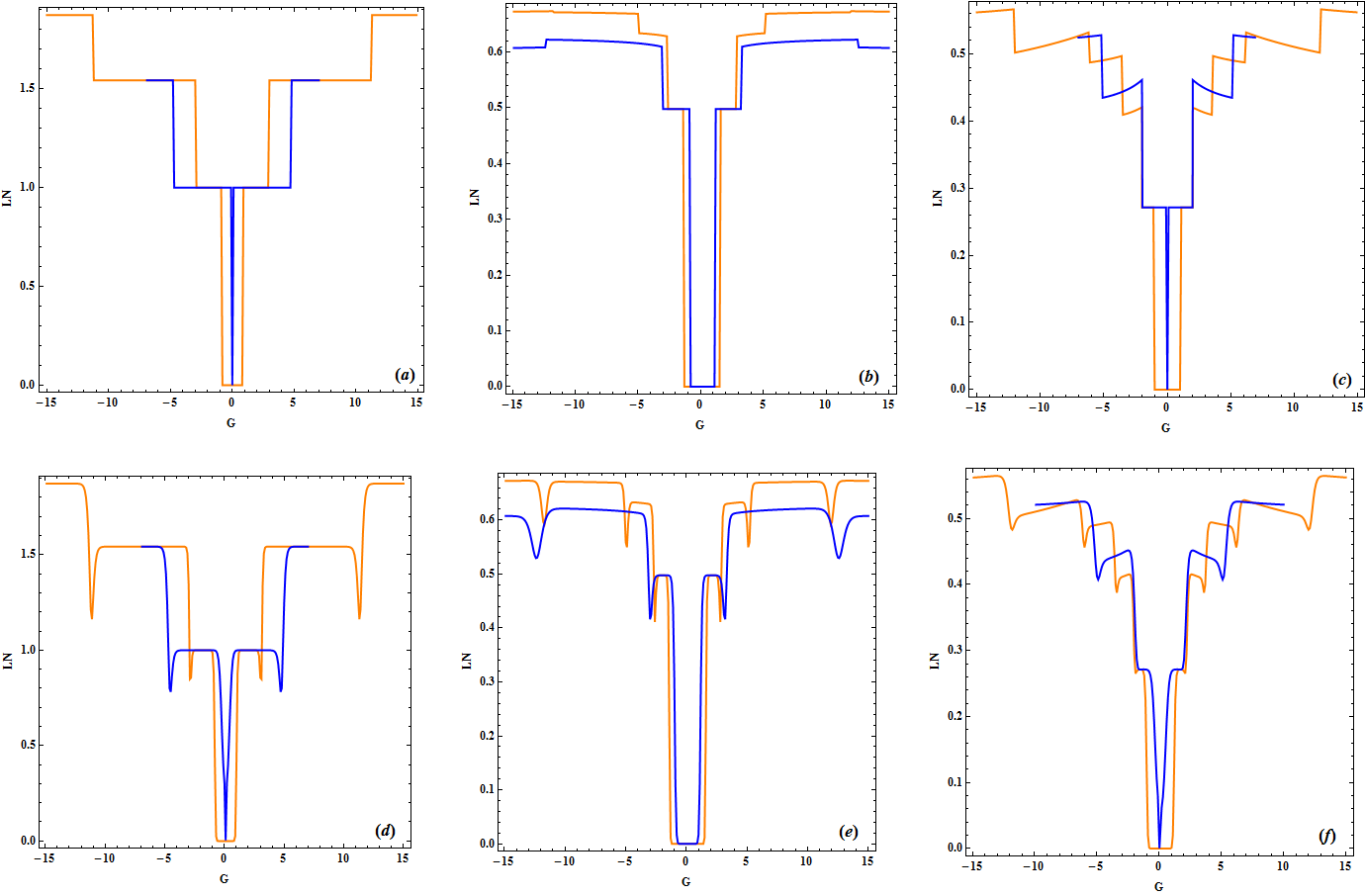}
	\caption{The logarithmic negativity $LN$ versus of DM interaction parameter $G$ }
	\label{ga}
\end{figure}

In  Fig.\ \ref{ga} we present the  behavior of logarithmic negativity $LN$ via DM interaction parameter $G$  at fixed $J = 1$, $D = 1$ for spin $1$ at $B = 5$ and for spin $\frac{3}{2}$ at $B = 9$. We fixed these values of $B$ near the $B_c$ for spin $1$ and spin $\frac{3}{2}$ respectively.

In Fig.\ \ref{ba} and Fig.\ \ref{ga} (a, b, c) the upper curves show the logarithmic negativity of $LN$ for spin $\frac{3}{2}$, and the lower curves for spin $1$ ($N = 2, 3, 4$). The same in plots (d, e, f) corresponds to a fixed temperature $k_B T = 0.1$.


\begin{figure}[!htbp]
  	\includegraphics[width=120mm]{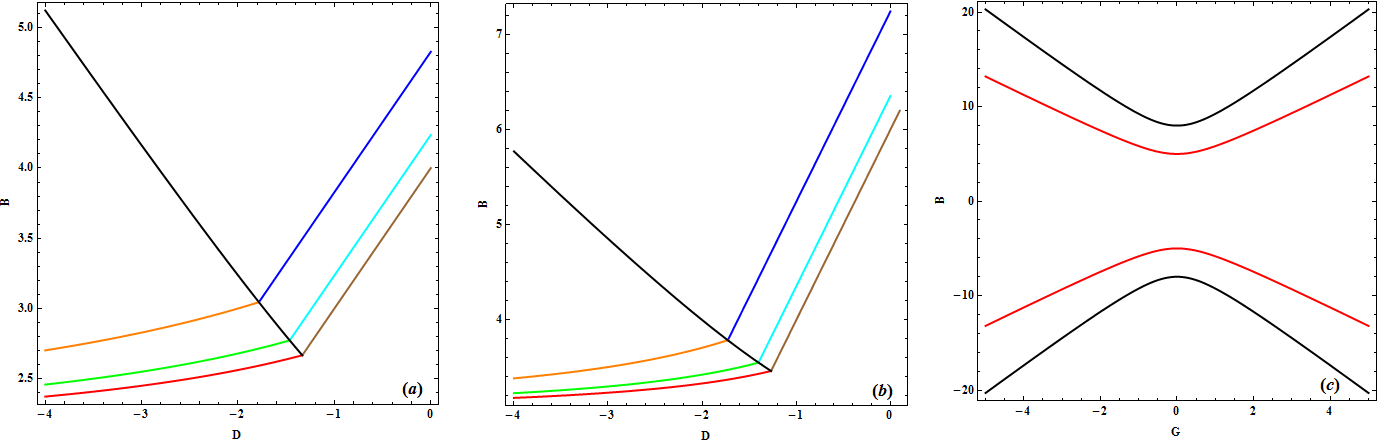}
	\caption{Ground-state phase diagrams in the $D - B$ plane at fixed value $J = 1$ and various
values of $G$ (from top to bottom $G = 1$, $G = \frac{1}{2}$, $G = 0$), plot (a) corresponds the spin $1$, plot (b) corresponds the spin $\frac{3}{2}$. Plot (c) corresponds the ground-state phase diagrams in the $G - B$ plane at fixed values $J = 1$ and $D = 1$. The entanglement area is between the upper and lower curves for the spin $1$. The region between middle curves corresponds the entanglement states, for the spin $\frac{3}{2}$.}
	\label{ph_dbg}
\end{figure}

We found out also the phase diagrams of ground-state in the $D - B$ and the $G - B$ planes at fixed value $J = 1$ for $N = 2$, see Fig.\ \ref{ph_dbg}. The entanglement region is located below the linear line and the adjacent curve for the corresponding values of the parameter $G$ (Fig.\ \ref{ph_dbg} (a) and (b)).
For example for spin $1$ and $N = 2$ at $G = 1$ equations of the linear line and the adjacent curve are obtained as follows:
\begin{equation}
B = 2 + 2\sqrt{2} + D \hskip 0.5cm \left(\text{for}\,\,D > -\frac{2}{7} \left(2 + 3\sqrt{2}\right)\right), \end{equation}
\begin{equation}
B = \frac{1}{2}\left(\sqrt{D^2 - 2D + 17} + D + 3\right) \hskip 0.5cm \left(\text{for}\,\,D < -\frac{2}{7} \left(2 + 3\sqrt{2}\right)\right). \label{p-d2}
\end{equation}
The curve connecting the corner points shows the possible values $B$ and $D$ of the triple points for various quantities of $G$ and in the case of spin $1$ and $N = 2$ is defined by:
\begin{equation}
B = 2\sqrt{\frac{1}{2} \left(D^2+\sqrt{(D+1)^2\left(D^2+1\right)}+D-1\right)+1}+D+2.
\end{equation}

The ground state of the system in the non entanglement region ($LN =0$) has a wave function corresponding to the state with all spins down ($\Psi_{GS} = |\downarrow\downarrow....\downarrow\rangle$).
The first nonzero plateau ground state wave function for the spin $1$ for chain sizes $N = 2$ and $N = 3$ is defined follow:
\begin{equation}
\Psi_{GS}(N = 2) =\frac{1}{\sqrt 2} (|\downarrow 0\rangle + (\cos(\frac{\pi }{4}) + i\cos(\frac{\pi }{4}))|0\downarrow\rangle)
\end{equation}
\begin{equation}
\Psi_{GS}(N = 3) =\frac{1}{\sqrt 3} (|0\downarrow\downarrow\rangle + i|\downarrow0\downarrow\rangle - |\downarrow\downarrow0\rangle)
\end{equation}
The coefficients for the spin $\frac{3}{2}$ are the same with replacements in the labels of vector of state $\downarrow$ by  $-\frac{3}{2}$  and $0$  by  $-\frac{1}{2}$.

\begin{figure}[!ht]
  	\includegraphics[width=120mm]{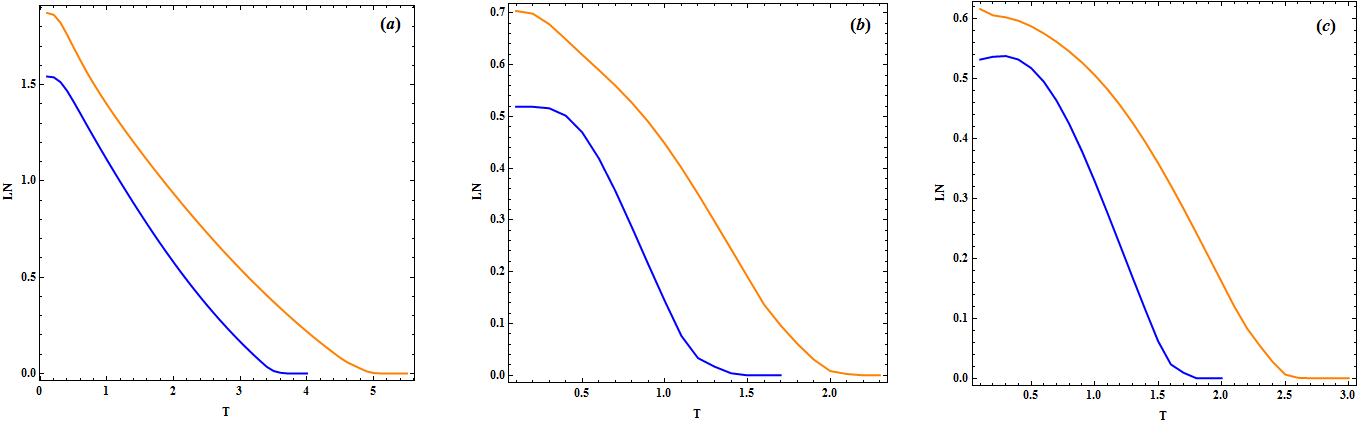}
	\caption{The upper curves represent the behaviour of  entanglement amount $LN$ for the spin $\frac{3}{2}$, and lowest curves show the same for the spin $1$. Plots (a), (b), (c) correspond $ N = 2$, $N = 3$, $N = 4$, respectively.}
	\label{t}
\end{figure}

The entanglement values $LN$ as a function of temperature $T$ (from now on $k_B = 1$) at fixed $J = 1$, DM interaction parameter $G = 1$ and single-ion anisotropy $D = 1$ are shown in Fig.\ \ref{t}

\begin{figure}[!htbp]
  	\includegraphics[width=80mm]{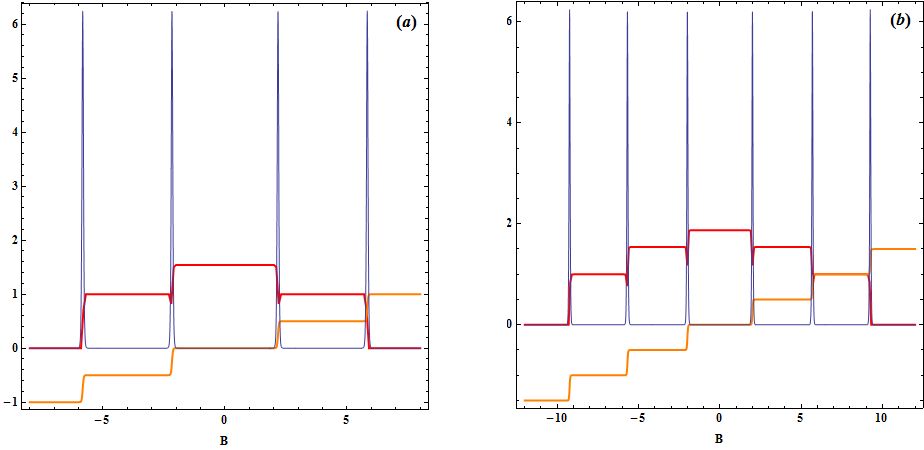}
	\caption{The logarithmic negativity (red curve), magnetization (orange curve) and magnetic susceptibility (blue curve) for $N = 2$ as a function of $B$ at $J = 1$, $D = 1$, $G = 1$ and $k_B T  = 0.05$. Plots (a) and (b) corresponds spin $1$ and $\frac{3}{2}$, respectively.}
	\label{k}
\end{figure}

The behavior of magnetization and magnetic susceptibility of the chain size $N = 2$ are presented in Fig.\ \ref{k}. As it was found, the plateau of magnetization and the plateau of logarithmic negativity coincide.
For chains of size $N = 3$ and $N = 4$, we get the results that are qualitatively similar.

N. A. acknowledges the receipt of the grant No. SCS 21AG-1C006.

\end{document}